\documentstyle[12pt]{article}
\def\be{\begin{equation}}
\def\ee{\end{equation}}
\def\bea{\begin{eqnarray}}
\def\eea{\end{eqnarray}}

\begin{document}

\begin{center}
{\Large{\bf Strings in the PP-wave Background from Membrane}}                  
										 
\vskip .5cm   
{\large Davoud  Kamani}
\vskip .1cm
 {\it Institute for Studies in Theoretical Physics and
Mathematics (IPM)
\\  P.O.Box: 19395-5531, Tehran, Iran}\\
{\sl E-mail: kamani@theory.ipm.ac.ir}
\\
\end{center}

\begin{abstract} 

In this paper we study strings with quantized masses in the pp-wave
background. We obtain these strings from the membrane theory. For
achieving to this, one of the membrane and one of the spacetime directions
will be identified and wrapped. From the action of strings in the
pp-wave background, we obtain its
mass dual action. Some properties of the closed and open strings in this
background will be studied.

\end{abstract} 
\vskip .5cm


\newpage

\section{Introduction}

It has been known that strings in the pp-wave NS backgrounds are exactly
solvable \cite{1}. The same is true for pp-waves on the {\it Ramond-Ramond}
backgrounds. Solvability in this context means that it is possible to
find explicitly the solutions of the classical string equations,
perform a canonical quantization and determine the Hamiltonian operator.
It was recently pointed out that
the plane wave metric supported by an $RR$ 5-form background \cite{2}
\bea
&~& ds^2 = 2dX^+dX^- -\mu^2\sum^8_{I=1}X^I X^I(dX^+)^2 + \sum^8_{I=1}
 dX^I dX^I,
\nonumber\\
&~& F_{+1234} = F_{+5678} = 2\mu,
\eea
provides examples of exactly solvable string models \cite{3}. That is,
the string action becomes quadratic in the light-cone gauge
$X^+ =x^+ +p^+ \tau$ \cite{1,3,4}.

Some properties of strings, propagating
in such plane-wave backgrounds have been previously investigated, in
particular in the Refs.\cite{3,5}.
This background is related, by a special limit \cite{6}, to the
$AdS_5 \times S^5$ background \cite{7}. The solvability of string theory in
this background has some common features with string theory on
$AdS_5 \times S^5$ \cite{8}. Another important application of string theory
in this background is that, it tests AdS/CFT correspondence beyond the
supergravity approximation \cite{9}.

In the other side we have the
membrane theory. It is characterized by the 11-dimensional spacetime
\cite{10}. For more review also see the Refs.\cite{11,12}. In fact,
the membrane theory, by compactification and double dimensional
reduction, provides an enlarged framework for the study of strings \cite{13}.
In this paper we study the connection between the membrane theory
and strings in the plane wave background.

Let one of the spatial directions of the membrane and of the spacetime be
identified and compactified on a circle.
In this case, the membrane can be viewed as infinite number of
strings in the pp-wave background with both positive and negative
quantized masses. This implies that the canonical quantization of
the membrane theory imposes the canonical quantization on the 
strings in the pp-wave background.
Therefore, the radius of the compactification will be fixed.

The action of these strings (i.e., $S$) describes interactions between the
worldsheets fields with opposite masses. By a transformation on the action
$S$, we obtain another useful action (i.e., $S'$), which contains
interactions between the worldsheets fields with the same masses.
Under some transformations on the
worldsheets fields, these actions are invariant or transform to each other.

We shall obtain the algebras and the Hamiltonians of the closed and open
strings. These Hamiltonians are normal ordered with respect to one of the
two indices that appear in each oscillator.

Note that we shall consider only the bosonic membrane. Therefore, we obtain
only bosonic strings. Generalization to the supersymmetric case is
straightforward. It is sufficient to consider supermembrane instead of
the bosonic membrane.
Since our formulation comes from the membrane theory, we shall
consider the eleven dimensional spacetime.

This paper is organized as the follows.
In section 2, we obtain the action of strings in the pp-wave background,
its quantization and its corresponding Hamiltonian from the
compactified membrane theory. In section 3, we
study the mass-dual theory and the symmetries of the actions $S$ and $S'$.
In section 4, we study the massive closed strings.
In section 5, the massive open strings will be analyzed.
\section{Massive strings from membrane}

Because of the non-linear structure of the membrane theory, we should
adopt some assumptions to simplify the problem. Therefore, let the
three- form gauge field of the 11-dimensional supergravity background vanish.
Furthermore, assume that the membrane propagates in the flat spacetime.
    
It is possible to choose a
gauge that identifies the eleventh dimension of the spacetime with the
third dimension of the membrane worldvolume. 
This gauge is $Z^{10} = \rho$, where $\rho, \sigma$ and
$\tau$ are coordinates of the membrane worldvolume \cite{12}.
This gauge also removes the cosmological constant
from the membrane action. By a light-cone type gauge \cite{12}, the
longitudinal coordinates $Z^{\pm} = \frac{1}{{\sqrt 2}}(Z^0 \pm Z^9)$
also can be removed from the action.
Therefore, the membrane action becomes
\bea
S=-\frac{T_2}{2}\int d^2 \sigma d \rho (\eta^{AB}\partial_AZ^I
\partial_BZ^I),
\eea
where $I \in \{1,2,...,8\}$. The metric of the membrane worldvolume is
$\eta_{AB}={\rm diag}(-1,1,1)$ and $T_2$ is the membrane tension.
The equation of motion of the membrane extracted from this action is
\bea       
(-\partial_\tau^2 + \partial_\sigma^2+\partial_\rho^2 )Z^I(\tau, \sigma,
\rho)=0.
\eea

If the $\rho$-direction of the membrane is wrapped around a circle
with the radius $R$, then
the $Z^{10}$-direction also is compact on the same circle. 
In fact, this is double dimensional
compactification, which compactifies both the spacetime and
the worldvolume on the same circle.
Upon dimensional reduction by shrinking the circle, the $Z^{10}$-direction
of the spacetime and the wrapped spatial dimension of the membrane will be
shrunk, for producing strings in ten dimension.

Now consider the set of functions
\bea           
\bigg{\{}\exp(\frac{in\rho}{R})\bigg{\}} \;\;\;,\;\;\; n\in Z\;,
\eea
where the range of $\rho$ is $0 \leq \rho \leq 2 \pi R$.
This set of functions forms a complete set. Therefore, we can expand
$Z^I(\tau, \sigma, \rho)$, in terms of them, i.e.,
\bea                        
Z^I(\tau,\sigma,\rho )=\sum^\infty_{n=-\infty} X^I_n(\tau,\sigma) q_n(\rho),
\eea
where $q_n(\rho)= a_n\exp(\frac{in\rho}{R})$.
Therefore, the membrane may be viewed as a ``tower of strings'' with the
coordinates $\{X^I_n(\tau , \sigma)| n \in Z\}$. As expected by
compactification, the coordinate $Z^I (\tau , \sigma,\rho)$
with respect to $\rho$ is periodic, i.e, $Z^I (\tau , \sigma,\rho+2\pi R)
=Z^I (\tau , \sigma,\rho)$. Since the membrane coordinate
$Z^I(\tau,\sigma,\rho )$ is real, we should have
\bea
&~& a^\dagger_n = a_{-n},
\nonumber\\
&~& X^{I\dagger}_n(\tau, \sigma)= X^I_{-n}(\tau,\sigma).
\eea

The membrane action (2) gives the
momentum conjugate to the coordinate $Z^I(\tau,\sigma,\rho)$ as
\bea
\Pi^I(\tau,\sigma,\rho )=T_2\sum_{n \in Z}\bigg{(}q_n(\rho)\partial_\tau
X^I_n(\tau,\sigma)\bigg{)}.
\eea
Quantization of the degrees of freedom is achieved by imposing the equal
time canonical commutation relation
\bea
\bigg{[}Z^I(\tau,\sigma,\rho )\; ,\; \Pi^J(\tau,\sigma',\rho' ) \bigg{]} 
=i \delta^{IJ} \delta(\sigma-\sigma')\delta(\rho-\rho').
\eea
In other words, it is
\bea
T_2 \sum_{n \in Z}\sum_{n' \in Z} \bigg{(}q_n(\rho)q_{n'}(\rho') 
\bigg{[}X^I_n(\tau,\sigma)\;,\;\partial_\tau X^J_{n'}(\tau,\sigma') \bigg{]}
\bigg{)}=i \delta^{IJ} \delta(\sigma-\sigma')\delta(\rho-\rho').
\eea
The delta functions in the right hand side imply that
\bea
\bigg{[}X_n^I(\tau,\sigma)\; ,\; \partial_\tau X_{n'}^J(\tau,\sigma') \bigg{]}
=\pi i \delta^{IJ} \delta(\sigma-\sigma') \delta_{n+n' , 0}.
\eea
The factor $\pi$ is introduced for later purposes. We shall see
that, this equation gives the quantization of the strings coordinates
$\{X^I_n(\tau , \sigma)\}$. Therefore, the equation (9) reduces to
\bea
\sum_{n \in Z} q_n(\rho) q_{-n}(\rho') =  \frac{1}{\pi T_2}
\delta(\rho-\rho').
\eea

Now consider the orthogonality equation
\bea
\int_0^{2\pi R} d \rho q_n(\rho)q_{n'}(\rho)
= 2\pi R a_n a_{-n}\delta_{n+n',0}.
\eea
According to this integral, the completeness of the functions
$\{q_n(\rho)\}$ gives
\bea
\frac{1}{2\pi R} \sum_{n \in Z}\bigg{(}\frac{1}{a_n a_{-n}} q_n(\rho)
q_{-n}(\rho')\bigg{)} = \delta(\rho-\rho').
\eea
Without lose of generality, we assume that $a_na_{-n} = 1$ for all values of
the integer $n$. In this case, the equations (11) and (13) give the relation
\bea
R = \frac{1}{2\pi^2 T_2}.
\eea
Therefore, 
the radius of compactification is fixed. That is, for the fixed parameter
$T_2$, we can not decompactify or shrink the $\rho$-direction,
to a circle with larger or smaller radius than $R$.

Finally, the action becomes
\bea
S=-\frac{1}{2\pi}\sum_{n\in Z} \int d\tau \int^{\pi}_0
d \sigma \bigg{(} \eta^{ab}
\partial_a X^I_n \partial_b X^I_{-n} + m^2_n X^I_n X^I_{-n}\bigg{)},
\eea
where the worldsheets metric is $\eta_{ab}= {\rm diag}(-1,1)$ and
\bea
m_n=\frac{|n|}{R} =|n|(2 \pi^2 T_2).
\eea
The action (15) can be splitted as $S= S_0 + \sum_{n\neq 0}S_n$.
The massless string has the action $S_0$ and the light-cone momentum
$p^+_0$. This part of the action
describes the usual string theory. Thus, $p^+_0$
is not zero. For the massive strings (i.e., $n \neq 0$) we have
$m_n= p^+_n \mu$. In fact, the mass $m_n$ is absolute value of
the momentum, conjugate to the
compact coordinate $\rho$, with the momentum number $n$.
The action (15) describes infinite number of strings that interact
with each other. Strings with the same mass couple to each other.
There is no coupling between the worldsheets fields with different masses.

Since the quantity $2\pi^2 T_2$
is quantum of the strings masses, one
interpretation is that, the membrane is a collection of infinite number
of strings that are sitting in the pp-wave background.
We can write $S= \sum_{n\in Z} S_n$, therefore,
each $S_n$ is a light-cone action, which
comes from the usual gauge fixing.
For this kind of gauge fixing see Refs.\cite{3,14}.

The equation of motion of a string with mass number $n$, extracted from
this action is
\bea
(-\partial^2_\tau + \partial^2_\sigma -m^2_n)X^I_n (\tau,\sigma)=0.
\eea
This equation also can be obtained from the equation (5) and the membrane
equation of motion (3).

The integral (12) and the equation (5) give the strings coordinates in terms
of the membrane coordinates
\bea
X^I_n (\tau , \sigma) = \pi T_2 \int_0^{2\pi R} d \rho
\bigg{(} q_{-n} (\rho) Z^I(\tau, \sigma, \rho)\bigg{)}.
\eea
Therefore, by an appropriate solution of the membrane, we can obtain 
solutions of the massive strings. But we shall obtain $X^I_n (\tau, \sigma)$
from the massive string equation (17).

{\bf The Hamiltonian}

The action (15) gives the momentum conjugate to the worldsheet field
$X^I_n(\tau , \sigma)$ as
\bea
\Pi^I_n(\tau , \sigma) = \frac{1}{\pi} \partial_\tau X^I_{-n}
(\tau , \sigma).
\eea
This implies that, strings with opposite mass numbers have conjugation
with each other. According to this conjugate momentum, the corresponding
light-cone Hamiltonian \cite{3} to the action (15) is 
\bea
H= \frac{1}{2\pi}\sum_{n\in Z}\bigg{[}\frac{1}{p^+_n}
\int^{\pi}_0 d \sigma
\bigg{(} \partial_\tau X^I_n\partial_\tau X^I_{-n} + \partial_\sigma X^I_n
\partial_\sigma X^I_{-n} + m^2_n X^I_n X^I_{-n} \bigg{)}\bigg{]}.
\eea
If we write $H= \sum_{n \in Z} H_n$, the condition (6) implies that each
$H_n$ (i.e., the Hamiltonian of the string with mass number $n$) is Hermitian.
This Hamiltonian also can be obtained from the membrane Hamiltonian.

According to the equation (10) we have the following canonical quantization
for the degrees of freedom of the action (15)
\bea
\bigg{[}X_n^I(\tau,\sigma)\; , \; \Pi^J_{n'}(\tau,\sigma') \bigg{]}
=i \delta^{IJ} \delta(\sigma-\sigma') \delta_{nn'}.
\eea
This quantization will be used for obtaining the algebra of the string modes.

The Hamiltonian (20) and the action (15) under the transformation
$X^I_n \rightarrow X^I_{-n}$, are invariant. In the next section we shall
study a general symmetry transformation of this Hamiltonian and the
action (15).
\section{Mass duality}

Under some transformations the action (15) remains invariant.
Furthermore, there are some other transformations that change this action to
another useful action. Consider the transformation
\bea
X^I_n \rightarrow Y^I_n = A^I_{(n)J}X^J_n + B^I_{(n)J}X^J_{-n},
\eea
where $A_n$ and $B_n$ are $8 \times 8$ matrices.
Now change $n \rightarrow -n$ and compare the result with the conjugation
of the above transformation (i.e., perform dagger on it), we obtain
\bea
&~& A^\dagger_n = A_{-n},
\nonumber\\
&~& B^\dagger_n = B_{-n}.
\eea
These also give $Y^{I\dagger}_n = Y^I_{-n}$. Note that this kind of
conjugation only acts
on the index $n$, therefore, spacetime indices under it are neutral. 

We now consider two cases. One case is that the action (15) and the
corresponding Hamiltonian (20) remain invariant and the other case is
that they transform to another useful action and Hamiltonian.

{\bf The invariance case}

For this case the matrices $A_n$ and $B_n$ satisfy the conditions
\bea
&~& A^T_n A_{-n} + B^T_{-n} B_n = {\bf 1},
\nonumber\\
&~& A^T_n B_{-n} + B^T_{-n} A_n = {\bf 0},
\eea
where $n$ is any integer number. 
According to these equations, these matrices also satisfy
the following equations
\bea
&~& A_n A^T_{-n} + B_n B^T_{-n} = {\bf 1},
\nonumber\\
&~& A_n B^T_n + B_n A^T_n = {\bf 0}.
\eea
Note that these equations are not independent of the conditions (24).

Define the matrix ${\cal{M}}_n$ as
\bea
{\cal{M}}_n= \left( \begin{array}{cc}
A_n & B_n\\
B_{-n} & A_{-n}                              
\end{array} \right).                              
\eea
Therefore, the equations (24) and (25) can be written as in the following
\bea
{\cal{M}}^T_n {\cal{M}}_{-n} = {\cal{M}}_n {\cal{M}}^T_{-n} = {\bf 1}.
\eea
These are symmetric conditions of the
action (15) and the Hamiltonian (20). For the zero mass case,
${\cal{M}}_0$ is an orthogonal matrix. 

{\bf The mass-dual action}

Let the matrix ${\cal{M}}_n$ satisfy the equations
\bea
{\cal{M}}^T_n {\cal{M}}_{-n} = {\cal{M}}_n {\cal{M}}^T_n = {\bf J},
\eea
where the matrix ${\bf J}$ has the definition
\bea
{\bf J}= \left( \begin{array}{cc}
{\bf 0} & {\bf 1}_{8 \times 8}\\
{\bf 1}_{8 \times 8} & {\bf 0}
\end{array} \right).                              
\eea
In this case, the action (15) and the Hamiltonian (20) transform to
the action
\bea
S'=-\frac{1}{2\pi}\sum_{n\in Z} \int d^2 \sigma \bigg{(} \eta^{ab}
\partial_a X^I_n \partial_b X^I_n + m^2_n X^I_n X^I_n\bigg{)},
\eea
and the Hamiltonian
\bea
H'=\frac{1}{2\pi}\sum_{n\in Z}\bigg{[}\frac{1}{p^+_n} \int^{\pi}_0 d \sigma
\bigg{(} \partial_\tau X^I_n \partial_\tau X^I_n +
\partial_\sigma X^I_n\partial_\sigma X^I_n + m^2_n X^I_n X^I_n\bigg{)}
\bigg{]}.
\eea
We call $S'$ as ``{\it mass-dual action }''  of the action $S$.
The action $S'$ implies that, for each mass number $n$
we have a light-cone action, which has been produced after the usual gauge
fixing \cite{3,14}. In other words, for each mass number ``$n$'', there
is a string in the pp-wave background.
According to this action, only strings with the same mass number
couple to each
other, while in the action $S$ strings with opposite mass numbers have
coupling.

The action $S'$ can be splitted to $S' = S'_0 + S'_+ +S'_-$,
where $S'_0$, $S'_+$ and $S'_-$ correspond to zero, positive and negative
masses, respectively. There is also $S'^\dagger_+ = S'_-$.
Similarly, the Hamiltonian $H'$ can be written as
$H' = H'_0 + \sum_{n=1}^\infty (H'_n + H'_{-n})$. The condition (6) implies
that each $H'_n$ (except $H'_0$) separately is not Hermitian. Therefore, we
should consider at least a system of two strings with opposite
mass numbers. This system has the Hamiltonian $H'_n + H'_{-n}$, which
is Hermitian.

According to the action $S'$, the momentum conjugate to the coordinate
$X^I_n$ is
\bea
{\Pi'}^I_n(\tau , \sigma) = \frac{1}{\pi} \partial_\tau
X^I_n(\tau , \sigma).
\eea
Therefore, the Hamiltonian $H'$ also can be extracted from the
action $S'$, as expected. Since $\Pi'^I_n = \Pi^I_{-n}$,
the quantization of the action
$S$, i.e., the equation (21), leads to the quantization of the action $S'$
as in the following
\bea
\bigg{[}X_n^I(\tau,\sigma)\; , \; \Pi'^J_{n'}(\tau,\sigma') \bigg{]}
=i \delta^{IJ} \delta(\sigma-\sigma') \delta_{n+n',0}.
\eea
This comes from the mass duality, that is, some properties of one theory
can be extracted from its dual theory.

Look at the usual string theory, that is, the worldsheet fields are massless.
Under the $T$-duality transformations we have the exchange
$X^I\leftrightarrow X'^I$ , where $X^I=X^I_L+X^I_R$ and $X'^I=X^I_L-X^I_R$.
Then, the string Hamiltonian is invariant. Furthermore, the conjugate momentum
$P^I = \frac{1}{\pi}\partial_\tau X^I$ and the momentum
$p^I = p^I_L+p^I_R$ of the string action,
exchange with the conjugate momentum
$P'^I = \frac{1}{\pi}\partial_\tau X'^I$ and the momentum
$2L^I= p^I_L-p^I_R$ of the $T$-dual action of string.
Similar to the coordinates $X^I$ and $X'^I$, we have $X^I_n$ and
$X^I_{-n}$. Under the exchange $X^I_n \leftrightarrow X^I_{-n}$, the
Hamiltonian (20) is invariant and the conjugate momenta $\Pi^I_n$
and $\Pi'^I_n$ change
to each other. We shall see that the momenta $p^I_n$ and $p'^I_n=p^I_{-n}$
also change to each other. Since we have the change $n \rightarrow -n$,
according to the
equation (16) the mass $m_n$ is invariant. For these reasons we
call the action (30) as the mass dual action of the action (15).

{\bf Transformations of the mass-dual action}

Application of the transformation (22) on the action $S'$ and on
the Hamiltonian
$H'$ leads to the two interesting cases. One case is that $S'$ and $H'$
remain invariant and the other case is that they transform to the action
$S$ and the Hamiltonian $H$.

For the case that $S'$ and $H'$ remain invariant, the matrix ${\cal{M}}_n$ is
orthogonal, i.e.,
\bea
{\cal{M}}^T_n {\cal{M}}_n = {\cal{M}}_n {\cal{M}}^T_n = {\bf 1}.
\eea
When $S'$ and $H'$ transform to the action $S$ and the Hamiltonian $H$,
the matrix ${\cal{M}}_n$ satisfies the equations
\bea
{\cal{M}}^T_n {\cal{M}}_n = {\cal{M}}_n {\cal{M}}^T_{-n} = {\bf J}.
\eea

Note that the matrices ${\cal{M}}_n$ and ${\bf J}$ satisfy the identity
\bea
{\cal{M}}_n {\bf J} = {\bf J} {\cal{M}}_{-n}.
\eea
Therefore, we can write the equations (28) and (35) in various forms.
This identity
also implies that $\det ({\cal{M}}_n) = \det ({\cal{M}}_{-n})$.

One may consider the action ${\bar S} = S+S'$. The equation of motion
extracted from the action ${\bar S}$ is the
equation (17). In the next sections we shall consider only the action $S$.
\section{Closed strings}

The solution of the equation of motion (17) for closed string is
\bea
X^I_n(\tau , \sigma)=x^I_n \cos(m_n \tau) + p^I_n
\frac{\sin(m_n \tau)}{m_n}
+i \sum_{l \neq 0} \frac{1}{\omega_{ln}}\bigg{(}
\alpha^I_{ln}e^{-i(\omega_{ln}\tau-2l\sigma)}
+{\tilde \alpha}^I_{ln}e^{-i(\omega_{ln}\tau+2l\sigma)}
\bigg{)},
\eea
where $\omega_{ln}$ is given by
\bea
\omega_{ln}= {\rm sgn}(l) \sqrt{(2l)^2 + m^2_n}.
\eea

The condition (6) implies that
\bea
&~& x^{I\dagger}_n = x^I_{-n},
\nonumber\\
&~& p^{I\dagger}_n = p^I_{-n},
\nonumber\\
&~& \alpha^{I\dagger}_{ln} = \alpha^I_{-l,-n},
\nonumber\\
&~& {\tilde \alpha}^{I\dagger}_{ln} = {\tilde \alpha}^I_{-l,-n},
\eea
where $l$ is non-zero integer and $n$ is any integer number. In other words,
conjugation changes the signs of both indices $l$ and $n$.
Note that $\alpha^I_{-l,-n}$ and ${\tilde \alpha}^I_{-l,-n}$
for positive $l$ and $n$ are creation operators, while
$\alpha^I_{ln}$ and ${\tilde \alpha}^I_{ln}$ are annihilation
operators. Furthermore,
$\alpha^I_{-l,n}$ and ${\tilde \alpha}^I_{-l,n}$
with respect to the index $l$ are creation operators, while with respect to
the index $n$ are
annihilation operators. Similar interpretation exists for the operators
$\alpha^I_{l,-n}$ and ${\tilde \alpha}^I_{l,-n}$.

The quantization (21) leads to the following commutation relations
\bea
&~&[x^I_n\; , \; p^J_{n'}] = i \delta^{IJ} \delta_{n+n',0},
\nonumber\\
&~& [\alpha^I_{ln}\; , \; \alpha^J_{l'n'}] = [{\tilde \alpha}^I_{ln}\; ,\;
{\tilde \alpha^J}_{l'n'}] = \frac{1}{2}\omega_{ln}
\delta^{IJ}\delta_{l+l',0}\delta_{n+n',0}.
\eea
For simplification consider
\bea
&~& \alpha^I_{ln} = \sqrt{\frac{|\omega_{ln}|}{2}}a^I_{ln},
\nonumber\\
&~& {\tilde \alpha}^I_{ln} =
\sqrt{\frac{|\omega_{ln}|}{2}}{\tilde a}^I_{ln}.
\eea
The operators $a^I_{ln}$ and ${\tilde a}^I_{ln}$ satisfy
\bea
&~& a^{I\dagger}_{ln} = a^I_{-l,-n},
\nonumber\\
&~& {\tilde a}^{I\dagger}_{ln} = {\tilde a}^I_{-l,-n}.
\eea
These operators have the algebra
\bea
[a^I_{ln} \; ,\; a^J_{l'n'}] = [{\tilde a}^I_{ln}\; , \;{\tilde a}^J_{l'n'}]
={\rm sgn}(l) \delta^{IJ}\delta_{l+l',0}\delta_{n+n',0}.
\eea

In terms of these oscillators, the Hamiltonian (20) for closed strings
can be written as $H=H_0+H_{\rm osc}$, where
\bea
H_{\rm osc} = \sum_{n \in Z} \sum_{l=1}^\infty
\bigg{(}\frac{1}{p^+_n} |\omega_{ln}|
(a^I_{-l,-n} a^I_{ln}+{\tilde a}^I_{-l,-n}{\tilde a}^I_{ln})\bigg{)} +
A+{\tilde A},
\eea
for the oscillating part. With respect to the index $l$ this part of
the Hamiltonian is normal ordered.
The constants $A$ and ${\tilde A}$ come from the algebra (43), and they are
\bea
A= {\tilde A} = 4\sum_{n \in Z} \sum_{l=1}^\infty\frac{|\omega_{ln}|}{p^+_n}=
-\frac{2}{3 p^+_0} +8\sum_{l=1}^\infty \sum_{n=1}^\infty
\frac{|\omega_{ln}|}{p^+_n}\;,
\eea
where we used $\sum_{l=1}^\infty l \rightarrow -\frac{1}{12}$.
According to the equation (42), for each string (i.e., for each $n$)
$H^{(n)}_{\rm osc}$ is Hermitian.

The zero mode part of the closed strings Hamiltonian has the form
\bea
H_0 = \frac{1}{2} \sum_{n\in Z}[\frac{1}{p^+_n} (p^I_n p^I_{-n}
+ m^2_n x^I_n x^I_{-n})].
\eea
The equations in (39) imply that for each string, $H^{(n)}_0$ also is
Hermitian. Let us define the operators $a^I_{0n}$ and $a^{I\dagger}_{0n}$ as
\bea
&~& a^I_{0n} = \frac{1}{\sqrt{2 m_n}}(p^I_n - i m_n x^I_n),
\nonumber\\
&~& a^{I\dagger}_{0n} = \frac{1}{\sqrt{2 m_n}}(p^I_{-n} + i m_n x^I_{-n}).
\eea
Therefore, we obtain
\bea
H_0 =\sum_{n \in Z}(\frac{1}{p^+_n}\omega_{0n}a^{I\dagger}_{0n}
a^I_{0n})+ A_0.
\eea
This form of $H_0$ is similar to $H_{\rm osc}$.
The equation (38) gives $\omega_{0n}= m_n$.
The constant $A_0$ only
depends on the radius of the compactification
\bea
A_0 = 4\sum_{n\in Z} \frac{\omega_{0n}}{p^+_n} = -4\mu\;,
\eea
where we used $\sum^\infty_{n=1} 1 \rightarrow -1$.
The operators in the zero mode part satisfy the equation
\bea
[a^I_{0n}\; , \; a^{J\dagger}_{0n'}] = \delta^{IJ}\delta_{nn'}.
\eea
In fact, for $n=n'$ this is the harmonic oscillator algebra.
\section{Open strings}

Variation of the action (15) gives the following boundary conditions for
the open string with the mass number $n$,
\bea
&~& (\partial_\sigma X^i_n)_{\sigma_0} = 0,\;\;\;\;
{\rm for\;the\;Neumann\;directions},
\nonumber\\
&~& (\partial_\tau X^a_n)_{\sigma_0} = 0,\;\;\;\;
{\rm for\;the\;Dirichlet\;directions},
\eea
where $\sigma_0 = 0\;, \pi$ shows the ends of the open string.
According to these conditions, the solutions of the equation of motion (17)
are
\bea
X^i_n(\tau , \sigma)=x^i_n \cos(m_n \tau) + p^i_n
\frac{\sin(m_n \tau)}{m_n}
+i\sum_{l \neq 0}\bigg{(} \frac{1}{\omega_{ln}}
\alpha^i_{ln}e^{-i\omega_{ln}\tau}
\cos (l\sigma)\bigg{)},  
\eea
\bea
X^a_n(\tau , \sigma)=
\sum_{l \neq 0}\bigg{(} \frac{1}{\omega_{ln}}
\alpha^a_{ln}e^{-i\omega_{ln}\tau}
\sin (l\sigma)\bigg{)},  
\eea
where $\omega_{ln}$ has definition
\bea
\omega_{ln}={\rm sgn}(l)\sqrt{l^2+ m^2_n}.
\eea
Note that this is different from $\omega_{ln}$ of the closed
string, that has been given by the equation (38).
The modes of open string also satisfy the equations
\bea
&~& x^{i\dagger}_n = x^i_{-n},
\nonumber\\
&~& p^{i\dagger}_n = p^i_{-n},
\nonumber\\
&~&\alpha^{I\dagger}_{ln} = \alpha^I_{-l,-n},\;\;\;\;I\in \{i,a\}.
\eea

The commutation relations of the open string oscillators
also can be obtained from the canonical
quantization (21), i.e.,
\bea
&~& [x^i_n \; , \; p^j_{n'}] = i \delta^{ij} \delta_{n+n',0},
\nonumber\\
&~& [a^I_{ln} \; , \; a^J_{l'n'}] =
{\rm sgn}(l) \delta^{IJ}\delta_{l+l',0}\delta_{n+n',0}\;\;\;\;,\;\;\;\;
I,J \in \{i,a\},
\eea
where we have $\alpha^I_{ln} = \sqrt{|\omega_{ln}|} a^I_{ln}$.

The Hamiltonian of the system is $H=H_0 + H_{\rm osc}$, where
\bea
H_0 = \frac{1}{2} \sum_{n\in Z}[\frac{1}{p^+_n}
(p^i_np^i_{-n} + m^2_n x^i_n x^i_{-n})],
\eea
for the zero mode part. Using the definitions in (47), $H_0$ can be written
as
\bea
H_0 = -\frac{1}{2}p\mu + \sum_{n\in Z} (\frac{1}{p^+_n}
\omega_{0n} a^{i \dagger}_{0n}a^i_{0n}),
\eea
where $\omega_{0n} = m_n $. Also $p$ shows the number of the Neumann
directions.
If all directions obey the Neumann boundary condition, there is $p=8$. In
this case, $H_0$ for the closed and open strings is the same, as expected.

For the oscillating part we have
\bea
H_{\rm osc} = \frac{1}{2} \sum_{n\in Z} \sum_{l \neq 0}
\bigg{(}\frac{1}{p^+_n}|\omega_{ln}| a^I_{-l,-n}a^I_{ln}\bigg{)}.
\eea
Normal ordered form of this part with respect to the index $l$ is
\bea
H_{\rm osc} = C + \sum_{n\in Z} \sum_{l=1}^\infty \bigg{(}\frac{1}{p^+_n}
|\omega_{ln}|a^I_{-l,-n}a^I_{ln} \bigg{)},
\eea
where the constant $C$ is
\bea
C= - \frac{1}{3p^+_0} + 8 \sum_{l=1}^\infty \sum_{n=1}^\infty 
\frac{|\omega_{ln}|}{p^+_n}\;.
\eea
The Hamiltonians (58) and (60) imply that,
for each $n$ (i.e., for each string) $H^{(n)}_0$ and
$H^{(n)}_{\rm osc}$, separately are Hermitian.
\section{Conclusions}
By compactifying the bosonic membrane we obtained the action of
infinite number of interacting
strings (i.e., the action $S$) in the pp-wave background.
The radius of the compactification is fixed and has expression
in terms of the membrane tension.

We found some transformations that the above action remains invariant,
or transforms to another action, which was called ``mass dual
action''. The corresponding Hamiltonian (i.e., $H$) also transformed as like
as the action $S$. Under some other transformations the mass dual
action and its corresponding Hamiltonian remain invariant or change to
the action $S$ and the Hamiltonian $H$.

We obtained quantization of the massive strings from the
membrane quantization.
This implies that the quantized masses are proportional to the
membrane tension. By the mass-duality, the
quantization of the mass dual action $S'$ was obtained from the quantization
of the initial action $S$.

The modes of both closed and open strings depend on the mass number
``$n$''. Under the conjugation of the massive string coordinates,
the mass index of the modes changes its sign,
i.e., $n \rightarrow -n$.
This implies that a string oscillator with respect to this index can
be creation or annihilation operator.
Therefore, in the algebras, strings modes with opposite
but equal absolute values of the mass indices, have conjugation with each
other.

The Hamiltonians of both closed and open strings
in terms of the strings modes have been obtained. 
These Hamiltonians with respect to one of the indices of the
oscillators are normal ordered.
As expected, for each string the zero mode part and
the oscillating part of the Hamiltonian are Hermitian.

{\bf Acknowledgment}:

I am grateful to the referee of the Physics Letters B, for useful
comments, which helped me to improve the manuscript.


\end{document}